\documentstyle[preprint,aps]{revtex}
\begin{document}
\tighten
\draft
\title{ \ Pseudo-spin of time-like lepton and the solar neutrino problem}
\author{ Vo Van Thuan }
\address{ Institute for Nuclear Science and Technique\\
          Hoang Quoc Viet Street, Nghia Do, Hanoi, Vietnam\\ 
          E-mail: vkhkthn@netnam.org.vn }
\maketitle
\begin{abstract}

{\small Based on the dual principle in super-luminous Lorentz transformation
this work shows that{} pseudo-spin of a time-like bradyon appears to
space-like observers as iso-spin of a corresponding tachyon. Due to the weak
interaction, lepton-tachyon appears as neutrino with hidden imaginary
transcendent mass, suppressed by a factor of }$\rho \sim G_Fm_0^2$%
{\small \ compared to the rest mass }$m_0${\small \ of a corresponding
space-like lepton. Assuming a coexistence of tachyon dark matter in the
solar system, we show that the solar neutrino deficit might be explained by
depolarisation of time-like electrons due to elastic scattering in time-like
plasma and the maximal deficit would reach 0.5. Neither day-night nor
seasonal effects are expected due to dynamic balance of fluxes with opposite
pseudo-spin polarisation.\smallskip }

\end{abstract}

\section{\protect\ Introduction}

E. Recami et al. [1] suggested that space-like tachyon is , in fact,
time-like bradyon, travelling in three-dimension time and one-dimension
space. Space-like and time-like bradyons behave in their own ''spaces'', in
complete symmetry. However, space-like tachyon has never been observed in
practice. We have suggested in Refs. [2,3] that tachyon transcendent mass is
complex, which would be observed by means of the weak interaction.
Consequently, both minus square mass and real part of transcendent mass as
dynamic parameters should be suppressed by a factor of $\rho ^2\sim $ $%
G_F^2m_0^4$ compared to the rest mass $m_0$ ($G_F$ is Fermi coupling
constant). In case of lepton, the factor $\rho $ is negligible small and
lepton-tachyons are always identified as luxons, i.e. luminous particles
moving with speed of light, such as neutrinos. Based on this assumption we
have given a qualitative interpretation of the results of the oscillation
experiments at accelerators (including LSND, CARMEN and NOMAD) and nuclear
reactors [3]. As a further development, the present paper introduces an
alternative approach to solve the solar neutrino problem discovered 31 years
ago by R. Davis et al. [4] and reconfirmed in later experiments [5,6,7,8].
Within this approach one is able also to give a qualitative explanation to
the negative result of recent Super-Kamiokande experiments which have not
seen the day-night effect [9].
\section{\protect\ Transcendent mass of neutrinos as realistic tachyons}

As suggested in Refs. [2,3] the complex transcendent mass $m^\infty $ of
tachyon has a dominant imaginary part Im ($m^\infty $) and a minor real part
Re ($m^\infty $). For lepton-tachyons, we assume that the imaginary part may
show up only in an interference between the weak interaction and the
electro-magnetic or strong interaction, i.e. Im ($m^\infty $) = $\alpha
_i\rho m_0$, where $\alpha _i$ is the coupling constant of \textit{i}-type
interfering interaction. Absolute value of the minus square mass and the
real part are to be observed in full weak interaction and should be of the
second order of the Fermi constant as $\vert$Im ($m^\infty $)$\vert$$^2$ $%
=\alpha _i{}^2\rho ^2m_0^2$ =$\alpha _i{}^2m_0\mathit{\Gamma }$/2 and Re ($%
m^\infty $) = $\rho ^2m_0$ = $\mathit{\Gamma }$/2, where $m_0$ is the rest
mass and $\mathit{\Gamma =}$ 192$^{-1}\pi ^{-3}$G$_F^2m_0^5,$ the decay
width of a corresponding lepton-bradyon. Observable transcendent masses of
lepton-tachyons are given in Table 1.

\bigskip

\begin{center}
Table 1. Estimated observable transcendent mass of lepton-tachyons

\bigskip

\begin{tabular}{cccc}
\cline{1-1}\cline{2-2}\cline{3-3}\cline{4-4}
lepton-bradyon & electron & $\mu $-meson & $\tau $-lepton \\ 
\cline{1-1}\cline{2-2}\cline{3-3}\cline{4-4}
m$_0$, MeV & 0.51 & 105.6 & 1780 \\ 
$\mathit{\Gamma }$, eV & $< 5.10^{-22}$ (*) & 2.5 10$^{-10}$ & \symbol{126}10%
$^{-3}$ \\ 
$\rho ^2$ & 5.10$^{-28}$ & 1.2 10$^{-18}$ & 3. 10$^{-13}$ \\ 
Im(m$^{\propto })\alpha _i^{-1}$, eV & 1.1 10$^{-8}$ & 0.116 & 979 \\ 
$\vert$Im(m$^{\propto })|^2\alpha _i^{-2}$, eV$^2$ & 1.2 10$^{-16} $ & 1.34
10$^{-2}$ & 9.5 10$^5$ \\ 
Re(m$^{\propto })$,eV & $\leq $2.5 10$^{-22}$ & 1.25 10$^{-10}$ & 5. 10$%
^{-4} $ \\ \hline
\end{tabular}

\bigskip

(*) \textit{A similar formula for} $\mathit{\Gamma }$ \textit{of unstable
leptons was extended to electron}
\end{center}

\smallskip These lepton-tachyons are supposed to exist as neutrinos. As a
result, neutrinos should behave in their own super-luminous reference frame
as time-like bradyons, which, according to the dual principle, should have
similar properties as space-like bradyons in sub-luminous reference frame.
Particularly, they have ''spin'', which, in difference from the space-like
leptons, is a dipole moment in three-dimension time. To a space-like
observer, this pseudo-spin looks like an iso-spin.

\section{\protect\ An alternative solution to the solar neutrino problem%
}

The solar neutrino fluxes measured in different experiments are reviewed in
Table 2.

\bigskip

\begin{center}
Table 2. Solar neutrino flux (SNU): measurements and theory

\medskip

\begin{tabular}{cccc}
\hline
Exp. \& Method & Theor. calculation & Exp. & Exp./Theor. \\ \hline
\multicolumn{1}{l}{Homestake:} & \multicolumn{1}{l}{} & \multicolumn{1}{l}{}
& \multicolumn{1}{l}{} \\ 
\multicolumn{1}{r}{Chlorine;total[5]} & 9.3$+1.2-1.4$ & 2.55$\pm 0.14\pm
0.14 $ & 0.27 \\ 
\multicolumn{1}{l}{Kamiokande:} &  &  &  \\ 
\multicolumn{1}{r}{water;B$^8$[6]} & 6.6$\pm 0.9$ & 2.805$\pm 0.19\pm 0.33$
& 0.42 \\ 
\multicolumn{1}{l}{Sage:} &  &  &  \\ 
\multicolumn{1}{r}{Gallium;total[7]} & 137+8-7 & 74$\pm 12+8-7$ & 0.54 \\ 
\multicolumn{1}{l}{Gallex:} &  &  &  \\ 
\multicolumn{1}{r}{Gallium;total[8]} & 137+8-7 & 77$\pm 9+4-5$ & 0.55 \\ 
\hline
\end{tabular}

\bigskip
\end{center}

In Table 2 the measured neutrino flux is significantly smaller compared to
the calculation from the standard solar model (SSM) [10]. The data of Ref.
[5] is by a factor of 3 below prediction, while other data are more or less
equal half of prediction [6,7,8]. The average ratio of total neutrino
spectrum measurements in Refs. [5,7,8] is 0.46 $\pm $0.06. All efforts to
improve the theoretical SSM have not succeeded to explain the experiments. A
natural way was to assume that probably there is oscillation of solar
neutrinos, which might be enhanced by MSW resonance mechanism in matter [11]
or in vacuum [12]. However, Super-Kamiokande collaboration showed that there
is no day-night effect which would happen if the resonance enhancement took
place in the Earth matter [9]. The seasonal variation, if exists, would be
less than 5\% of the solar neutrino flux observed by Super-Kamiokande [13].
For solar neutrino the oscillation parameter $\Delta m^2$ $\sim $10$%
^{-5}$-10$^{-6}$ eV$^2$ is too small to agree with oscillation from
atmospheric or accelerator data.

This situation stimulated us to look for a new alternative solution of the
problem, considering neutrino as tachyon in our sub-luminous reference
frame, or as time-like bradyon in super-luminous frame.

V. De Sabbata et al. [14] proposed that a black-hole may be a trap for
tachyons, in which, i.e., inside the horizon they behave them-selves as
time-like bradyons. As a black hole is involved in gravitation, its radius R
calculated by Schwarzshild model has a linear dependence on mass: R (km)$%
\approx $GM$_b$ /c$^2$ (G is the gravitation constant). For example, a black
hole of the same radius as of the Sun should have mass M$_b$ = 3.10$^5$ M$_s$
(where M$_s$ is solar mass). Those black holes could not exist inside the
Sun, because of huge gravitational collapse they might cause to the Sun as a
middle-size star. The picture will change, as according to our hypothesis,
the black holes contain tachyon matter, which hardly interacts with
space-like one. The real part of tachyon mass Re(\textrm{m}\textit{)}
involved into gravitation is suppressed by a weak interaction factor of $%
\rho ^2$=192$^{-1}\pi ^{-3}$G$_F^2$m$^4$. For the black hole containing
time-like mass with an abundance as in the Sun, we may put m $\approx $1.2 m$%
_p$ (where m$_p$ is proton mass), which leads to the observable mass of the
black hole M$_b^{*}$ $\approx $ 10$^{-8}$M$_s$. This means that the tachyon
black hole of extremely huge transcendent mass is able to exist inside the
Sun without collapse!

Consequently, we assume, that there can be in the solar center a
concentration of tachyon dark matter of time-like plasma, containing
time-like electrons and lightest time-like nuclei, i.e. proton and helium.
As a result, solar neutrinos have to pass through this area, scattering in a
similar way as space-like electrons interact with plasma. In according to
Ref. [15] we considered a 100\% initially polarised electron flux and
calculated its depolarisation, which depends on the thickness of scattering
plasma medium as shown in Fig.1.

\medskip

\medskip \bigskip

In the Fig.1, we see that passing in matter thickness of more than 60 kg.cm$%
^{-2}$ (equivalent to $\approx $400m through normal solar matter), electrons
loose the polarisation completely. As a result, 50\% of particles are
oriented toward their moving direction, and 50\% - backward. Concerning the
flux intensity, we suggest that, in difference from normal solid matter,
plasma does not capture electrons, but elastically scatter them only. In a
significant thickness of homogenous plasma the scattering is always
isotropic and the electron beam intensity must be conserved.

Further we are going to extend this formalism to the flux of solar
neutrinos, considering them as time-like electrons or positrons, initially
polarised 100\% by alignment of their pseudo-spin toward the time direction,
i.e. toward the future. Passing through tachyon dark matter plasma, confined
in the black hole, neutrinos are elastically scattered and their pseudo-spin
after depolarisation is equally oriented toward and backward the time axis,
and their beam intensity is not changed due to absence of capture. Among
them, 50\% remaining oriented towards the future direction, i.e. co-existing
with us as space-like observers, and can be observed as solar neutrinos.
While 50\% are moving to the past, i.e. becoming anti-particles. For
definition we recall that Dirac neutrino is space-like left-handed, while
Dirac anti-neutrino is right-handed and they are different each from other.
In case of solar neutrino, its anti-particle after pseudo-spin reversion
remains the same space-like left-handed. Therefore, solar neutrino and its
anti-particle are identical time-like bradyons having opposite pseudo-spin
projection (to- and anti-time axis), which are completely similar to a
normal electron able to exist in one of the two states with opposite spin
projection +1/2 or -1/2 along the space-like moving direction. As a result,
we found that anti-particle of neutrino is not Dirac one, but it looks like
Majorana one. Moreover, time-reversed neutrino are not real anti-particle
from the view-point of time-like observers. Meanwhile, it is left-handed, in
difference from realistic observable Dirac anti-neutrino. Going backward to
the past time, Majorana anti-neutrinos, consisting 50\% of the total initial
solar neutrino flux, are not able to be detected by terrestrial
observatories, and they remain sterile neutrinos. In our terrestrial
experiments one can observe only 50\% neutrinos going toward the future in
consistency with experimental data shown in Table 2. This is one of
alternative explanation of the observed total rate of solar neutrinos.

Similar to a normal space-like unpolarised electron beam, the total solar
neutrino flux reached the Earth consists of two components with opposite
pseudo-spin projections. These two components equally interact with tachyon
dark matter in the media they are passing through from the Sun to the Earth.
A dynamic balance is established to keep the ratio 50:50 between the
observable neutrino flux and the sterile component unchanged. This behaviour
predicts the absence of both day-night and seasonal effects.
Super-Kamiokande experimental data [9,13] seem to confirm this assumption.

\section{\protect\ Conclusion}

In this paper we show that neutrinos are realistic tachyons and according to
the dual principle, are time-like leptons, simultaneously, which have
pseudo-spin instead of conventional space-like spin. Due to weak
interaction, the observable minus square mass and real part of transcendent
mass of those tachyons should be dynamically suppressed by a factor $\rho ^2$%
, which is negligible small to measure and neutrinos are detected almost as
luminous particles. Suppression of tachyon mass leads to a possibility for a
black hole to exist inside the Sun, which serves a location of dark matter
plasma. As a result, after depolarisation of neutrinos by reversing their
pseudo-spin while passing through the dark matter area, only half of them
are remaining detectable. The other 50\% become some kinds of sterile
neutrinos, moving backward to the past. Those sterile neutrinos are
recognised as Majorana anti-neutrinos. A dynamic balance between observable
and sterile components in interaction with tachyon dark matter on the way
from the Sun to the Earth might explain the absence of the day-night and
seasonal effects. Thus, the solar neutrino problem might be solved in this
way.

The author appreciates the financial support of the National Basic Research
Program for Natural Science sponsored by the Ministry of Science, Technology
and Environment of Vietnam for 1999.

\bigskip

\begin{figure}
\caption{ Depolarisation of a polarised electron beam vs. thickness of
scattering medium }
\label{f1}
\end{figure}

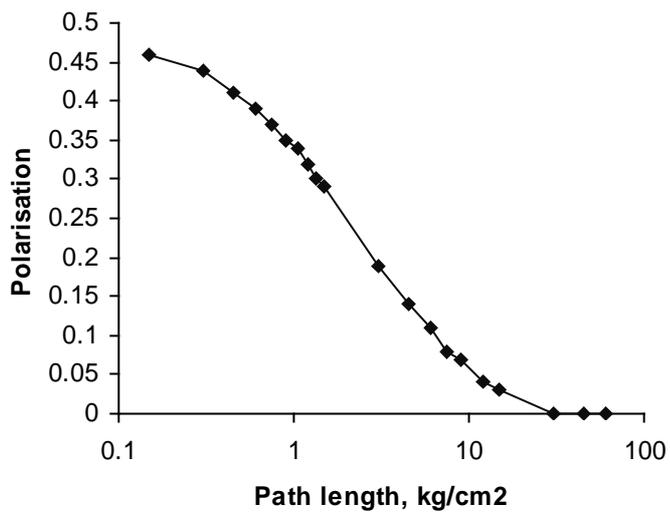
\end{document}